\documentclass[aps,prb,twocolumn,groupedaddress,amsmath,amssymb]{revtex4}

\usepackage{graphicx}
\usepackage{dcolumn}
\usepackage{bm}
\usepackage{color}
\usepackage{subfigure}

\newcommand{\ybco}{YBa$_2$Cu$_3$O$_{6+\delta}$}
\newcommand{\lbco}{La$_{2-x}$Ba$_x$CuO$_4$}
\newcommand{\lbcoo}{La$_{15/8}$Ba$_{1/8}$CuO$_4$}
\newcommand{\lsco}{La$_{2-x}$Sr$_x$CuO$_4$}

\newcommand{\lnsco}{La$_{2-x-y}$Nd$_y$Sr$_x$CuO$_4$}
\newcommand{\bscco}{Bi$_2$Sr$_2$CaCu$_2$O$_{8+\delta}$}
\newcommand{\ccoc}{Ca$_{2-x}$Na$_x$CuO$_2$Cl$_2$}

\newcommand{\Tc}{T_{\rm c}}

\newcommand{\Qch}{{\vec Q}_{\rm c}}

\newcommand{\Qsp}{{\vec Q}_{\rm s}}

\newcommand{\Qp}{{\vec Q}_{\rm p}}

\providecommand{\up}{\ensuremath{\uparrow}} 
\providecommand{\down}{\ensuremath{\downarrow}} 

\begin{document}

\title{Inter-layer Josephson coupling in stripe-ordered superconducting cuprates}

\author{Alexander Wollny}
\author{Matthias Vojta}
\affiliation{Institut f\"ur Theoretische Physik, Universit\"at zu K\"oln, Z\"ulpicher Stra{\ss}e 77, 50937 K\"oln, Germany}

\date{July 28, 2009}


\begin{abstract}
Motivated by experiments on \lbco\ which suggest that stripe order co-exists
with two-dimensional pairing without inter-layer phase coherence over an extended
range of temperatures, we determine the inter-layer Josephson coupling in the presence of
stripe order. We employ a mean-field description of bond-centered stripes, with
a zero-momentum superconducting condensate and alternating stripe directions pinned by the
low-temperature tetragonal (LTT) lattice structure.
We show that the Fermi-surface reconstruction arising from strong stripe order can
suppress the Josephson coupling between adjacent layers by more than an order of magnitude.
\end{abstract}

\pacs{}
\maketitle


\section{Introduction}

Stripe order is a fascinating phenomenon in cuprate superconductors.\cite{stripe_rev1,stripe_rev2}
Originally detected in neutron-scattering experiments on \lnsco,\cite{jt95} this combination
of uni-directional spin and charge order was found in other members of the ``214''
family of cuprates as well.\cite{stripe_rev2}
Remarkably, incommensurate low-energy spin fluctuations, often interpreted
as precursors to stripe order, are seen not only in \lsco,
but also in \ybco\cite{hayden04,hinkov08a} and \bscco.\cite{xu09}
Together with STM measurements, which detected signatures of charge stripes (albeit with
substantial disorder) on the surface of \bscco\ and \ccoc,\cite{kohsaka07}
these findings suggest that the tendency toward stripe order is common to underdoped cuprates.

What is less clear is the role of stripes for superconductivity.
A large body of experiments appears consistent with the concept of competing
superconducting and magnetic order parameters,\cite{stripe_rev2}
including e.g. the magnetic-field enhancement of spin-density wave (SDW) order.
However, a few observations also point to a co-operative interplay of SDW and pairing.
In \lbcoo, the onset of SDW order upon decreasing temperature is accompanied by a significant
drop in the in-plane resistivity, while bulk Meissner effect sets in at much lower
temperatures.\cite{li07,jt08}
This intermediate-temperature regime of \lbcoo\ has been interpreted in terms of fluctuating
2d pairing, without inter-layer phase coherence.
(A related phenomenon is the suppression of the Josephson plasma resonance seen in the
optical-conductivity measurements of \lsco\ upon application of a moderate c-axis
field.\cite{schafgans})
In order to explain the absence of an effective inter-layer coupling, the existence of a
stripe-modulated (i.e. finite-momentum) superconducting condensate, a so-called pair density wave (PDW),
was postulated.\cite{berg07,berg08b}
Indeed, in the absence of a uniform condensate, the lowest-order Josephson coupling between
neighboring layers vanishes, if the condensate modulation direction alternates from layer
to layer -- the latter being the result of the in-plane lattice distortions inherent to the
LTT structure.
However, some properties of the PDW state are not easily compatible with
experiments: both photoemission and STM have established a $d$-wave like gap in the
stripe-ordered state above $\Tc$, whereas the PDW state has a full Fermi
surface.

In this paper, we shall investigate whether a primarily uniform condensate in a stripe-ordered state
could be compatible with the scenario of fluctuating 2d pairing in \lbcoo,
i.e., the absence of inter-layer phase coherence.
To this end, we calculate the inter-layer Josephson coupling for a mean-field model of
superconducting charge-density wave (CDW) state with realistic parameter values.
The stripe order induces a reconstruction of the Fermi surface, with rotation symmetry
breaking in each CuO$_2$ plane. As the stripe direction alternates from layer to layer,
the reconstructed Fermi surfaces of adjacent layers do not match in momentum space. This
effect leads to a significant suppression of the Josephson coupling; an additional
suppression arises from incommensurate antiferromagnetism of realistic amplitude.


\section{Order parameters and mean-field model}

We start by enumerating the relevant order parameters for a superconducting stripe state.
Charge and spin density waves, with wavevectors $\Qch$ and $\Qsp$,
are related to expectation values of particle-hole bilinears in the singlet and triplet
channel, respectively. For instance, a CDW is characterized by
non-zero $F_c(k) = \sum_\sigma \langle c_{\vec k+\Qch,\sigma}^\dagger c_{\vec k\sigma}\rangle$ where
$c_{\vec k\sigma}^\dagger$ creates an electron with momentum $\vec k$ and spin $\sigma$.
The superconducting condensate can have both a uniform component
and a modulated (PDW) component with wavevector $\Qp$, such that
$\langle c_{\vec k+\Qp,\uparrow} c_{\vec k\downarrow}\rangle\neq 0$.
On symmetry grounds, a collinear SDW will induce a CDW with wavevector $\Qch = 2\Qsp$,
a PDW will induce a CDW with $\Qch = 2\Qp$, and in the presence of a uniform condensate
a CDW will induce a PDW with $\Qp=\Qch$.

\begin{figure}[t]
\centerline{
\includegraphics[width=3.5in]{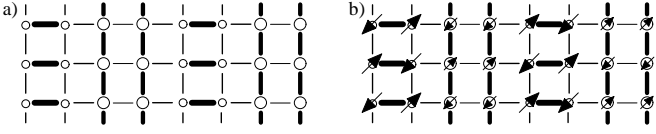}
}
\caption{
Schematic real-space structure of valence-bond stripes at doping 1/8,\cite{mvor08}
with circle sizes (line widths)
denoting on-site hole densities (bond strengths).
a) Paramagnetic state with dominant $d$-wave modulation.
b) Additional spin modulation with anti-phase domain walls
on hole-rich stripes.
}
\label{fig:cartoon}
\end{figure}

In our modelling, we start from two CuO$_2$ layers with homogeneous $d$-wave pairing and
then add CDW/SDW modulations.
Each layer $i\!=\!1,2$ is described by a quasi-particle model of electrons moving on a square
lattice, with the Hamiltonian
\begin{align}
{\cal H}^{i}
  &= \sum_{\vec k\sigma}(\varepsilon_{ \vec k} -\mu) c^\dagger_{\vec k i\sigma}c_{\vec k i\sigma}
   + {\cal H}_\text{DW}^{i} + {\cal H}_\text{P}^{i}
\label{full_Ham}
\end{align}
and the in-plane dispersion
$\varepsilon_{ \vec k} = - 2 t (\cos k_x +\cos k_y ) - 4 t' \cos k_x \cos k_y - 2 t'' (\cos 2 k_x +\cos 2 k_y)$
with $t'=-t/4$ and $t''=t/12$.

The symmetry-breaking orders are implemented non-selfconsistently at the mean-field level
by ${\cal H}_\text{DW}$ and ${\cal H}_\text{P}$. We restrict our attention to
bond-centered stripes of period 4 (8) in the charge (spin) sector which appear most
compatible with experiments at doping 1/8.\cite{stripe_rev1,stripe_rev2} The density-wave
part ${\cal H}_\text{DW}$ is given by
\begin{align}
{\cal H}_\text{DW}^{i}
 &= \sum_{\vec{k}\sigma}
  \Phi_c^i c^\dagger_{\vec k i \sigma} c_{\vec k + \Qch^i,i\sigma}
+ \Phi_s^i c^\dagger_{\vec k i \sigma} c_{\vec k + \Qsp^i,i\sigma} +  \text{h.c.}, \nonumber\\
 \Phi_c^1(\vec k\sigma) &= -e^{i\frac{\pi}{4}} \left(\cos(k_x+\frac{\pi}{4}) - \cos k_y\right) \delta t, \nonumber\\
 \Phi_s^1(\vec k\sigma) &=  - \sigma \frac{1}{\sqrt2}\frac{1+e^{-i\frac{\pi}{4}}}{1+\sqrt2}\delta\mu^{\sigma}
\label{CDWSDWOP}
\end{align}
where $\Qch^1 = (\pi/2,0)$, $\Qsp^1 = (3\pi/4,\pi)$, and the $\Phi_{c,s}^2$ are obtained from
$\Phi_{c,s}^1$ by $x\leftrightarrow y$. $\Phi_c$ implements a hopping modulation on the
bonds of strength $\delta t$, resulting in so-called valence-bond stripes.\cite{vs,mvor08}
Such bond-charge modulation, Fig.~\ref{fig:cartoon}a, is compatible with the STM data of
Ref.~\onlinecite{kohsaka07} and implies a strong $d$-wave component in the form factor $F_c$
of the CDW order parameter.
The associated collinear spin order, Fig.~\ref{fig:cartoon}b, is implemented via
a spin-dependent chemical potential $\delta\mu^{\sigma}$ in $\Phi_s$.\cite{millis07}

The pairing part ${\cal H}_\text{P}$ is dominated by a uniform $d_{x^2-y^2}$-wave pairing
mean field $\Delta_0$. In addition, a pairing modulation of amplitude $\delta\Delta$
is assumed along with the CDW,
resulting in a pattern of bond pairing amplitudes qualitatively similar to
Fig.~\ref{fig:cartoon}a, but with $d$-wave sign structure.
\begin{align}
{\cal H}_\text{P}^i
&= \sum_{\vec k} \Delta_{\vec k} c^\dagger_{\vec k i\up}c^\dagger_{-\vec k i\down}
   + \Phi_p^i   c^\dagger_{\vec k i\up} c^\dagger_{-\vec k-\Qch,i\down}\nonumber\\
   &+ \Phi_p^{i*} c^\dagger_{\vec k+\Qch^i,i\up} c^\dagger_{-\vec k i\down} + \text{h.c.}, \nonumber\\
 \Delta(\vec k)&= \Delta_0 \left(\cos k_x-\cos k_y\right),\nonumber\\
 \Phi_p^1(\vec k)&=-e^{i\frac{\pi}{4}}\frac{1}{2} \left(\cos(k_x+\frac{\pi}{4})+\cos
 k_y\right) \delta \Delta,
\label{PairOP}
\end{align}
and $\Phi_p^2$ is obtained from $\Phi_p^1$ by $x\leftrightarrow y$.


\section{Inter-layer Josephson coupling}

\subsection{Inter-layer tunneling}

The Hamiltonian of the full system of two adjacent CuO$_2$ layers is given by
\begin{align}
 {\cal H} = {\cal H}^1 + {\cal H}^2
 + \sum_{\vec k\sigma}
 \left[t_\perp(\vec k)c^\dagger_{\vec k 1\sigma}c_{\vec k 2 \sigma} + \text{h.c.}\right]
\label{Ham_coupled}
\end{align}
with ${\cal H}^i$ given in Eq.~\eqref{full_Ham}, and
the inter-layer hopping matrix element\cite{t_perp}
\begin{align} 
\label{tperp}
  t_\perp(\vec k) = \frac{t_\perp}{4}(\cos k_x-\cos k_y)^2.
\end{align}

To calculate the Josephson coupling it is convenient to multiply global
phase factors $\theta_i$ to the superconducting mean fields $\Delta_0$ and $\delta\Delta$
in layer $i$.
Then, the Josephson coupling measures the inter-plane phase stiffness:
\begin{align}
 J_J &= \frac{1}{2}\left[F(\delta\theta=\pi)-F(\delta\theta=0)\right]
\end{align}
where $F = F^1 + F^2 + \delta F (\delta\theta)$,
$F^i$ is the free energy of the isolated layer $i$,
$\delta F$ is the inter-layer tunneling contribution to the free energy,
and $\delta\theta=\theta_2-\theta_1$.

Assuming $t_\perp \ll t$, $\delta F$ can be determined in second-order perturbation theory
in $t_\perp$:
\begin{align}
 \delta F &= \frac{1}{\beta}\text{tr}\left( \hat{\mathcal G^1}\hat T \hat{\mathcal G^2} \hat T \right)
      = \frac{1}{N}\sum_{\vec{k}}  \delta F_{\vec k}, \nonumber \\
 \delta F_{\vec k} &=\frac{1}{\beta}\sum_{\omega_n} t_\bot(\vec{k})^2 \sum_{\alpha,\beta=0}^1\mathcal
   (-)^{\alpha+\beta} \mathcal G^{1,\alpha\beta}_{\vec k n}\mathcal
   G^{2,\beta\alpha}_{\vec k n}
\label{deltaf}
\end{align}
where $\mathcal G^i$ is the full Green's operator on layer $i$,
$\hat T$ is the inter-layer tunneling operator from $t_\perp$,
$\beta$ is the inverse temperature, and $N$ the number of unit cells.
The indices $\alpha,\beta$ denote particle-hole space. Since the stripe directions are orthogonal,
we have to consider the full (non-reduced) Brillouin zone (BZ).
For a period-4 CDW (period-8 CDW+SDW) in each superconducting layer,
the calculation of $\delta F$ involves the diagonalization of a $8\times8$ ($16\times16$)
Hamiltonian matrix to construct the Green's functions required in Eq.~\eqref{deltaf}.


\subsection{Results}

The numerical calculations have been performed at zero temperature, with parameters
$t=0.15$\,eV, $\Delta_0=0.024$\,eV, and fixed doping $x=1/8$. For the homogeneous case,
this corresponds to $\mu=-0.126$\,eV.

\begin{figure}[t]
\centering
\subfigure[][]{%
\includegraphics[scale=0.325]{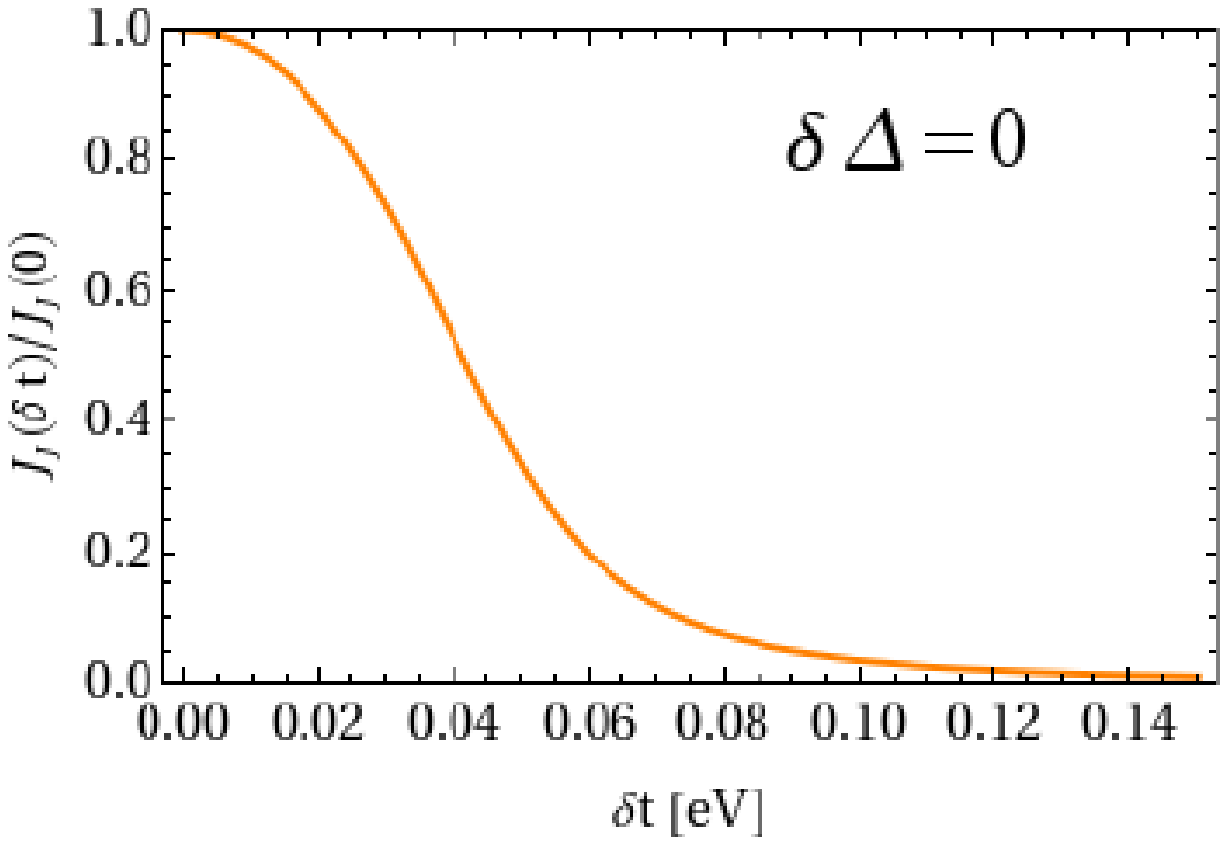}
}
\subfigure[][]{%
\includegraphics[scale=0.325]{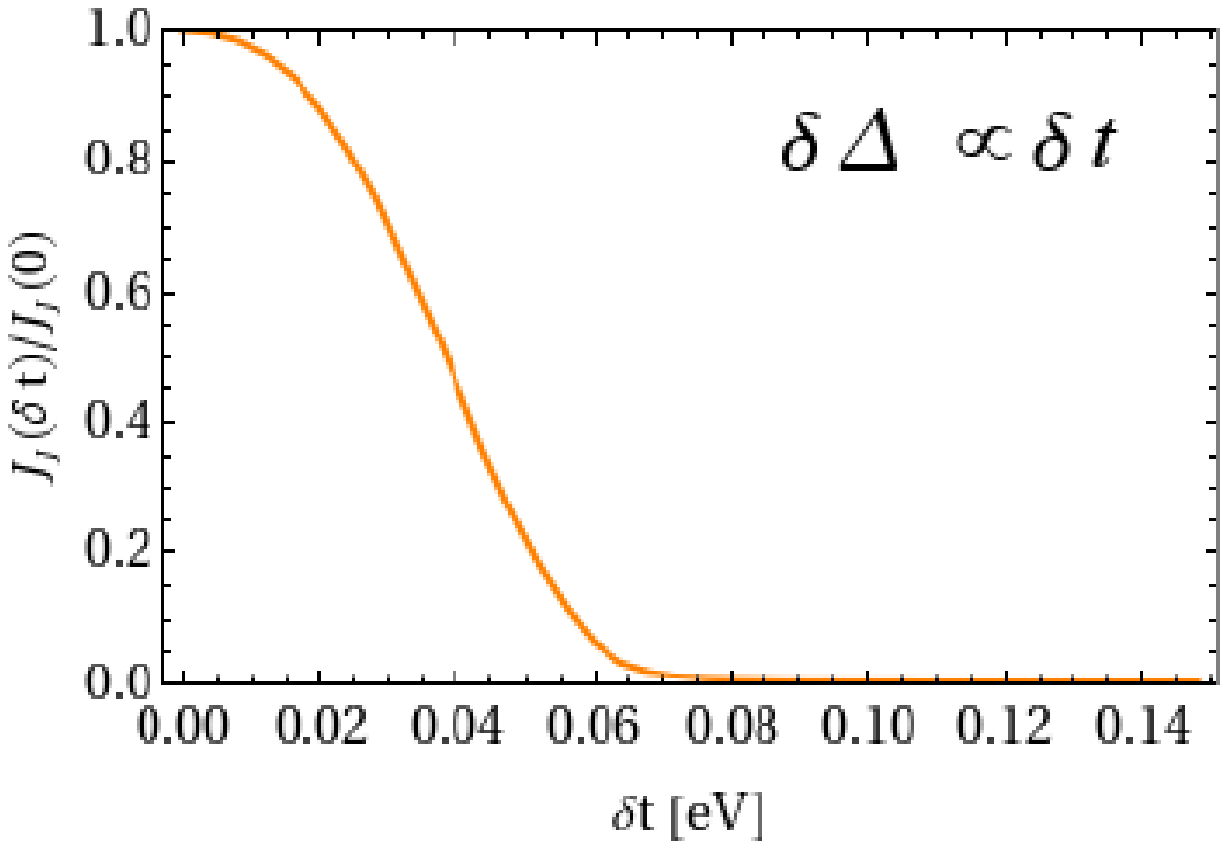}
}
\subfigure[][]{%
\includegraphics[scale=0.325]{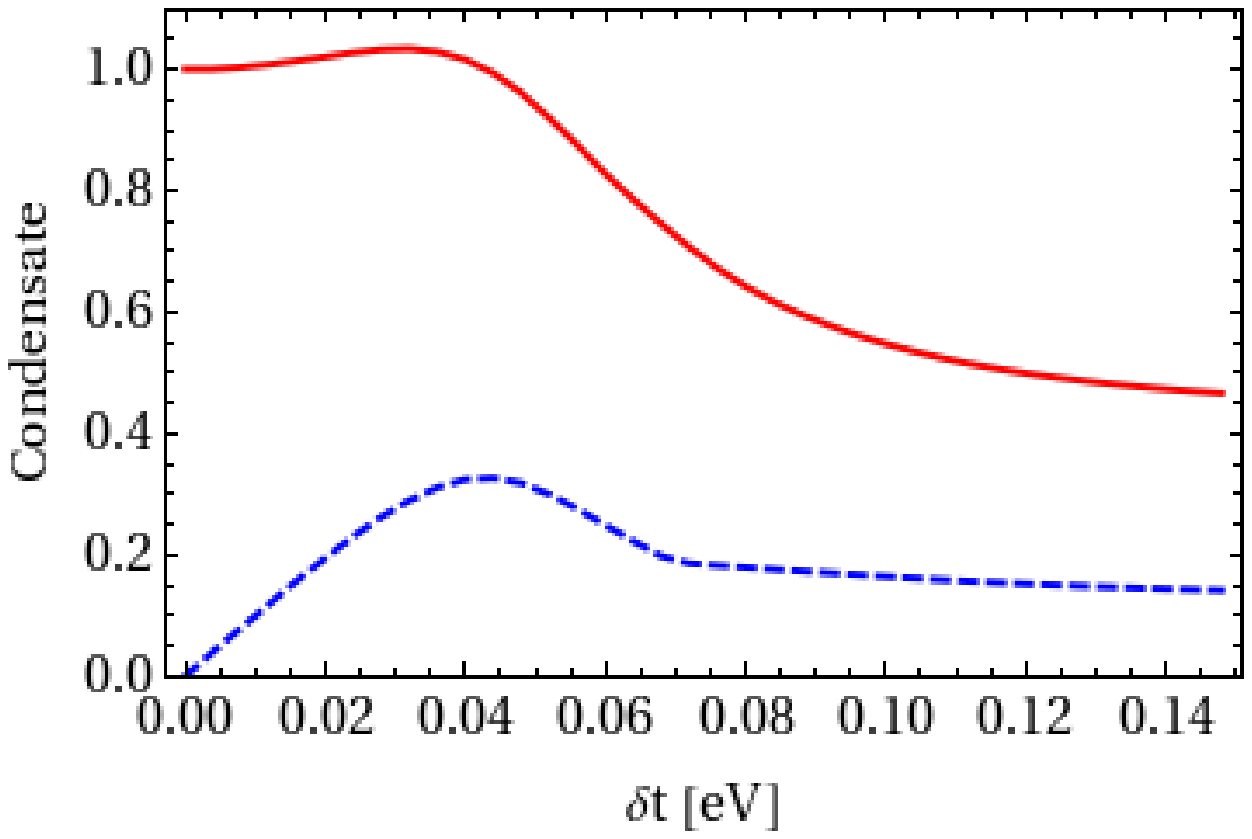}
}
\subfigure[][]{%
\includegraphics[scale=0.325]{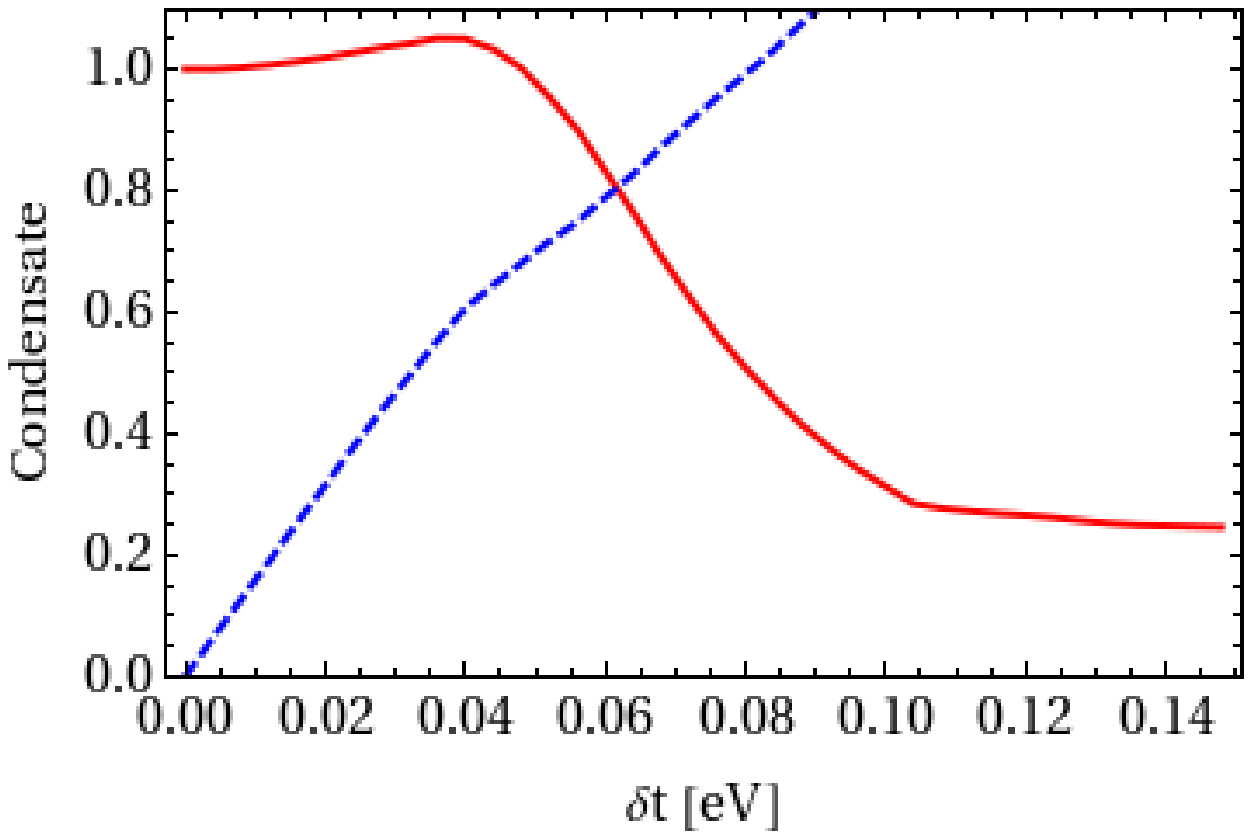}
}
\caption{
Inter-layer Josephson coupling $J_J$ (a,b)
and superconducting condensate amplitudes $\psi_0$ (solid)
and $\psi_{\Qch}$ (dashed) (c,d)
as functions of the hopping modulation strength $\delta t$,
at fixed doping $x=1/8$ and in the absence of magnetic order.
The modulation in the pairing field is $\delta\Delta/\delta t=0$ (a,c) and 1.3 (b,d).
The couplings and condensates are normalized w.r.t. the values
of $J_J$ and $\psi_0$ at $\delta t=\delta\Delta=0$.
}
\label{fig:Jcoupling_condensate}
\end{figure}

Results for $J_J$ for charge-only stripes as function of the hopping modulation $\delta t$
($\delta\Delta = \delta \mu^\sigma=0$) are shown in Fig.~\ref{fig:Jcoupling_condensate}a.
For large modulation amplitude, the Josephson coupling is seen to be strongly suppressed,
e.g. by roughly a factor 10 for $\delta t= 0.07$ eV.
A simultaneous modulation in the condensate mean field by $\delta\Delta$ (here chosen to
produce similar relative modulation strengths in the resulting bond kinetic energies and
pairings) suppresses the Josephson coupling even further,
Fig.~\ref{fig:Jcoupling_condensate}b.

Figs.~\ref{fig:Jcoupling_condensate}c and d show the corresponding evolution of the
homogeneous and modulated condensate amplitudes, $\psi_0$ and $\psi_{\Qch}$, calculated
from the solution of the mean-field Hamiltonian \eqref{full_Ham}.
Here, we define $\psi$ as the sum of the magnitudes of the $s$-wave (on-site), $s_{x^2+y^2}$-wave
and $d_{x^2-y^2}$-wave condensates calculated from the real-space pairing amplitudes
extracted from the solution of ${\cal H}^i$
(note that the $s$ and $d_{x^2-y^2}$ representations of the point group mix in the
presence of stripe order).
A modulated condensate $\psi_{\Qch}$ is always present for $\delta t\neq 0$,
but remains small if $\delta\Delta=0$.
In contrast, for $\delta\Delta\propto\delta t$ as in
Fig.~\ref{fig:Jcoupling_condensate}d, $\psi_{\Qch}$ increases and eventually dominates
over $\psi_0$.
A comparison between the evolution of $J_J$ and $\psi_0$ reveals that in the range of
$\delta t$ where $J_J$ drops dramatically, the uniform condensate $\psi_0$ displays a much
weaker depletion. This is also true for panels b and d where, at $\delta t \approx 0.08$\,eV,
$J_J$ is reduced by a factor $200$ while $\psi_0$ still has half of its original value.
From this we conclude that the primary source of the suppression of the Josephson coupling
in our calculation
is different from that of the PDW proposal by Berg {\em et al.}\cite{berg08b}
where the layer decoupling is due to the {\em absence} of a homogeneous condensate.

\begin{figure}[b]
 \centering
\subfigure[][]{
\includegraphics[scale=0.3]{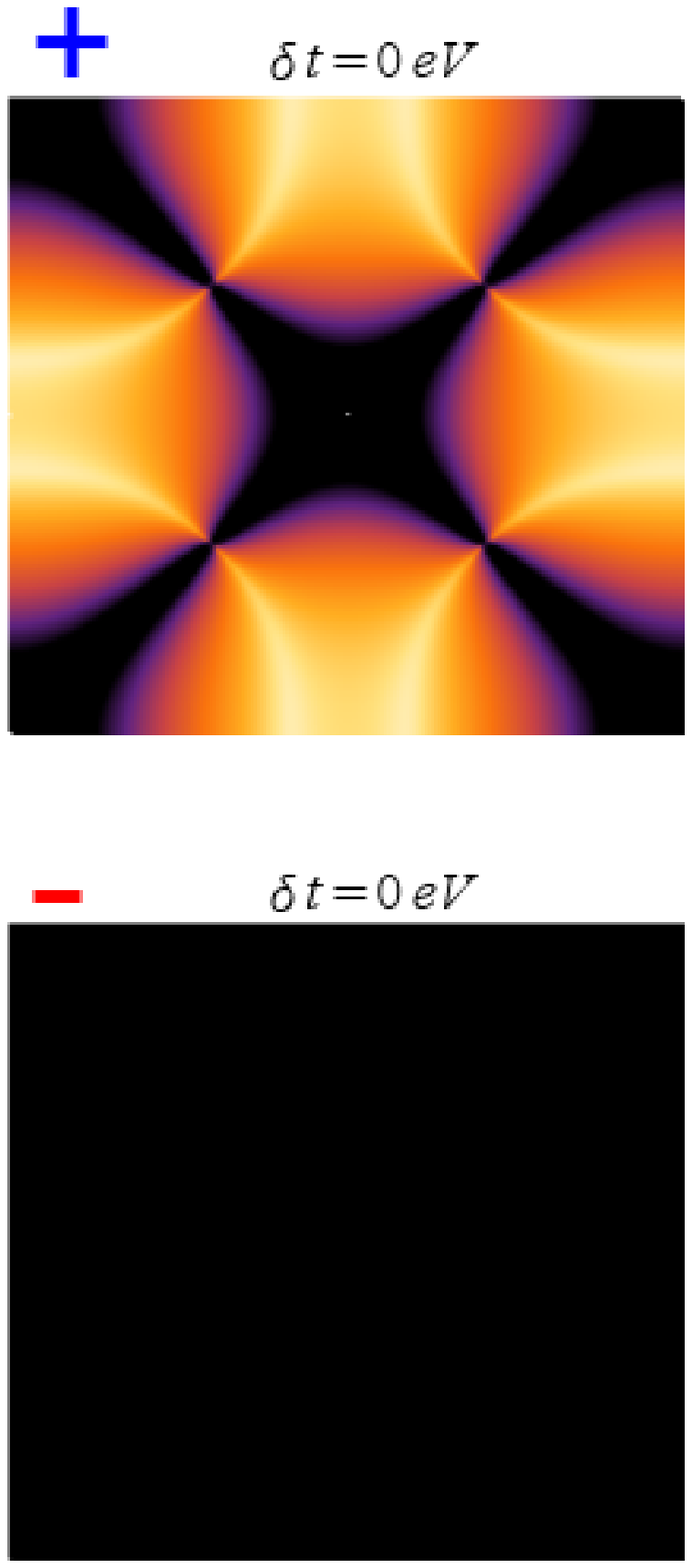}
}
\subfigure[][]{
\includegraphics[scale=0.3]{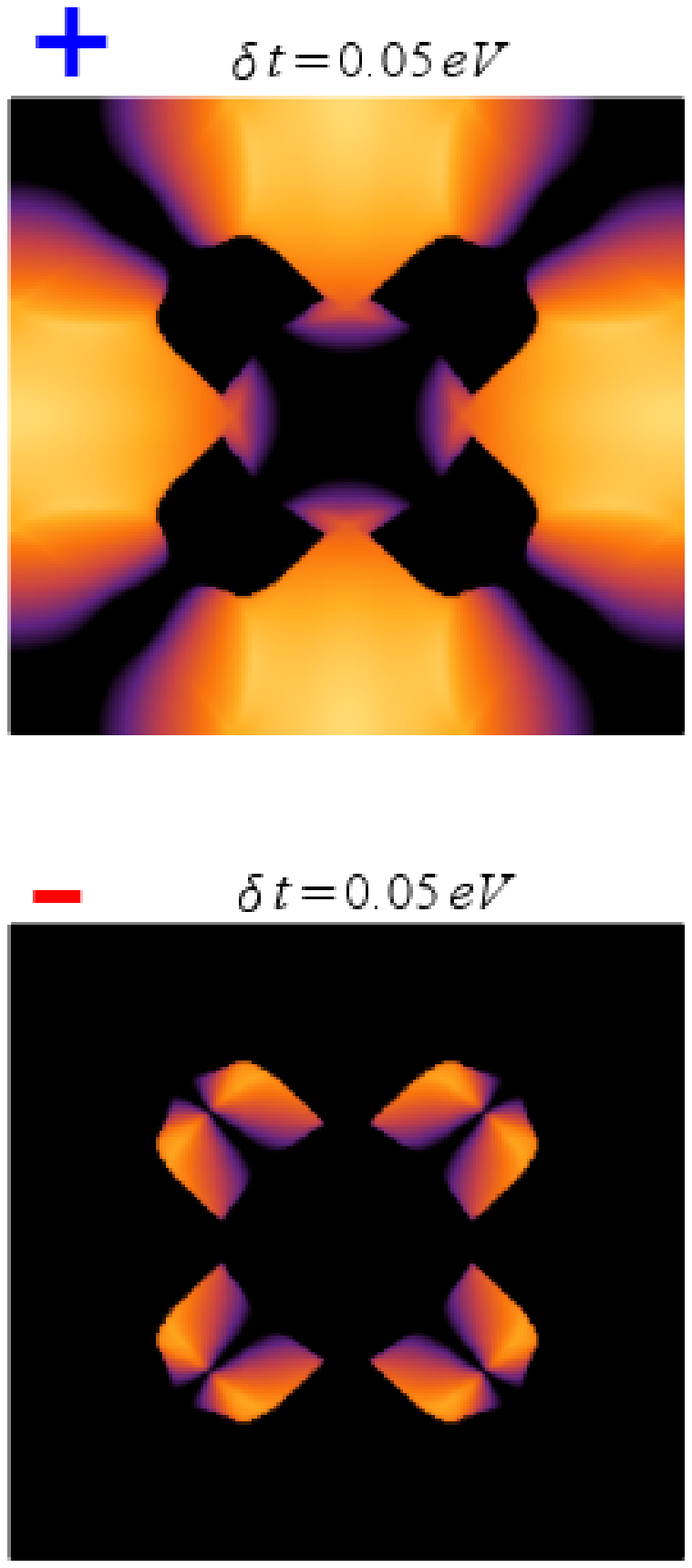}
}
\subfigure[][]{
\includegraphics[scale=0.3]{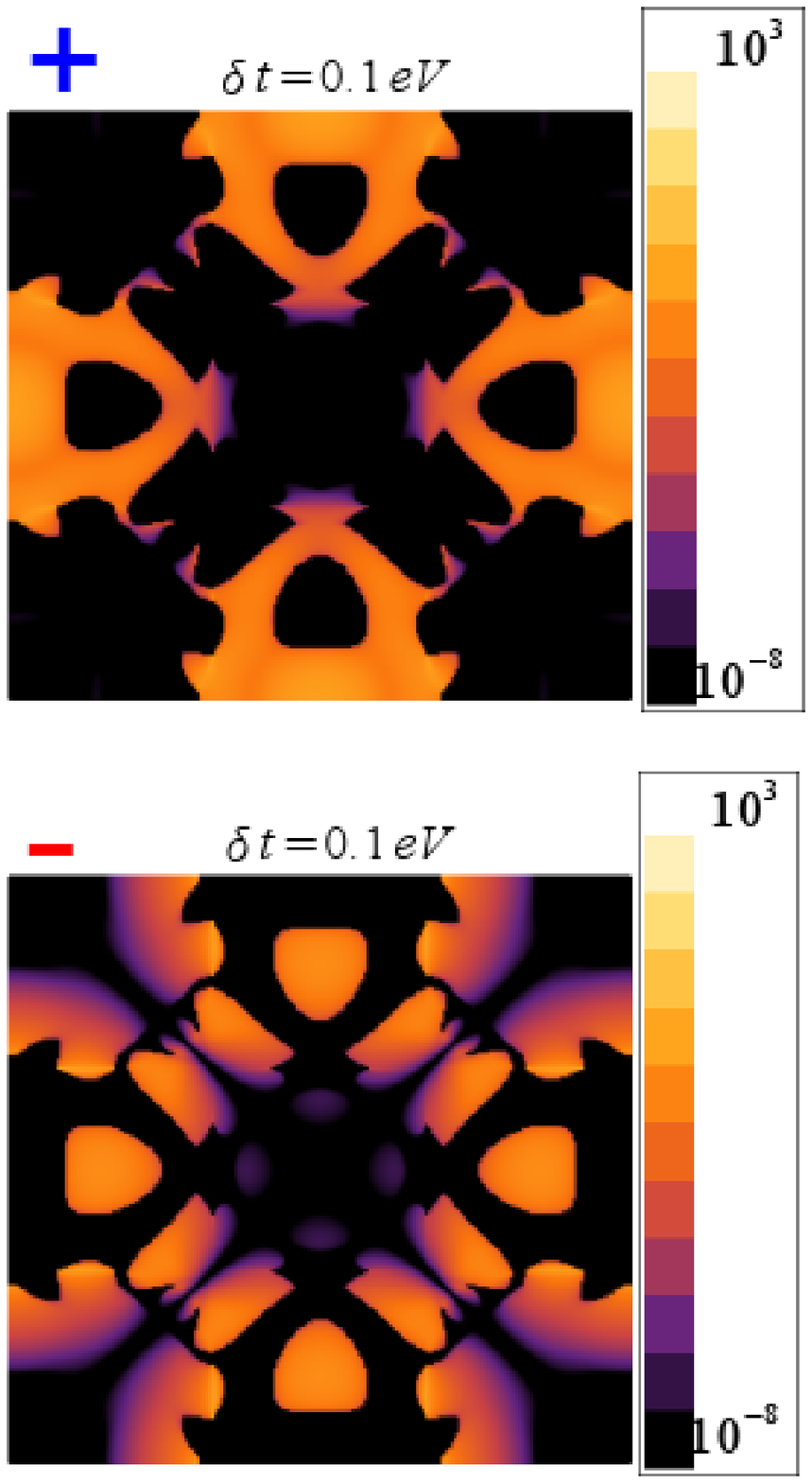}
}
\caption{
Positive (top) and negative (bottom) momentum-resolved contributions to $J_J$.
Each panel shows $|J_J(\vec k)/t^2_\perp|$ (see text) as function of $\vec k$
on a logarithmic intensity scale.
The modulation strength $\delta t$ is zero in a) and increases from b) to c).
The other parameters are as in Fig.~\ref{fig:Jcoupling_condensate}b,d.
From a) to c) the positive contributions near the antinodal points are reduced,
moreover negative contributions appear.
Note that the momentum dependence of the inter-layer tunneling \eqref{tperp}
suppresses the contributions near the diagonals.
}
\label{fig:Cstripes_J_plusminus}
\end{figure}

\begin{figure}[t]
\centerline{
\subfigure[][]{%
\includegraphics[scale=0.35]{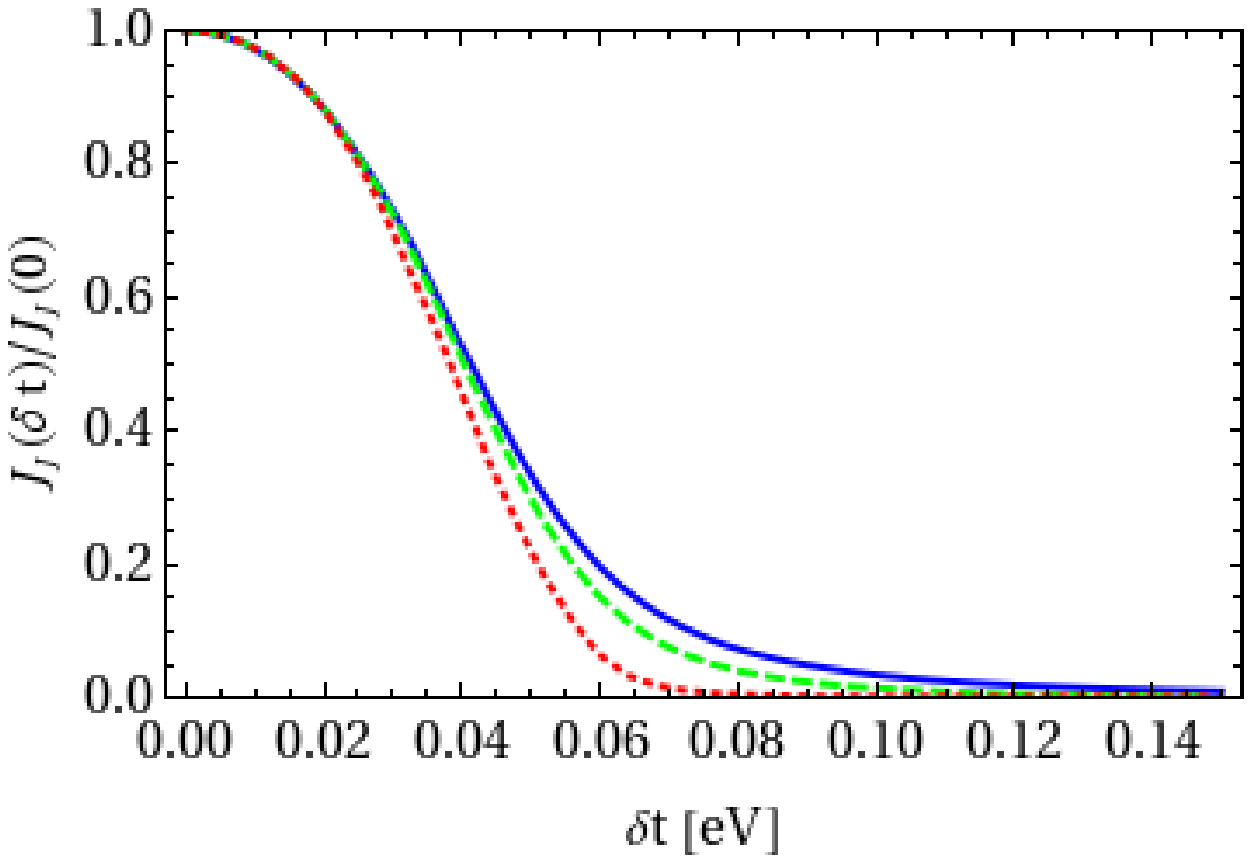}
}
\subfigure[][]{%
\includegraphics[scale=0.35]{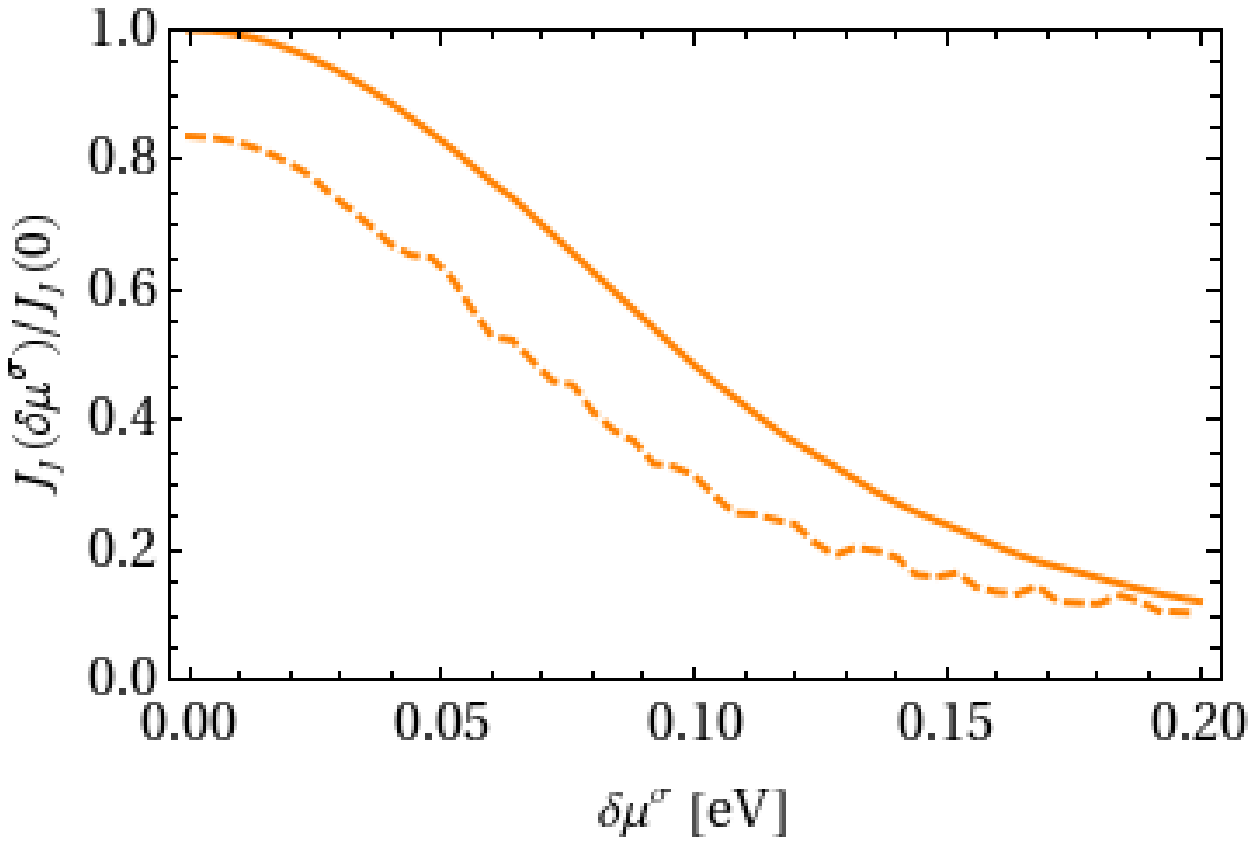}
}
}
\caption{
Normalized inter-layer Josephson coupling $J_J$ at doping 1/8 for
a) paramagnetic stripes with hopping modulation $\delta t$ and accompanying pairing
modulations $\delta\Delta/\delta t=0$ (solid), 0.5 (dashed), 1.3 (dotted), and
b) antiferromagnetic stripes with magnetic modulation set by $\delta\mu^\sigma$,
for $\delta t=\delta\Delta=0$ (solid) and
for moderate fixed kinetic energy modulation of $\approx\pm25\%$ (dashed)
achieved via adjusting $\delta t$ and keeping $\delta\Delta=1.3\delta t$.
}
\label{fig:coupling_CDWdiffratios_SDW}
\end{figure}

Analyzing our results further, we identify the momentum-space mismatch of the Fermi surfaces of the two layers,
arising from the orthogonal stripe modulation, as the main source of the suppression of $J_J$.
This mismatch is also accompanied by a mismatch of the nodal lines of the superconducting
order parameter in the two layers, due to broken rotational symmetry in each layer.
These effects can be nicely seen in the momentum-resolved contributions
$J_J(\vec k) = \delta F_{\vec k}(\delta\theta\!=\!\pi)-\delta F_{\vec k}(\delta\theta\!=\!0)$
to the Josephson coupling, shown in Fig.~\ref{fig:Cstripes_J_plusminus}.
In the homogeneous case, the largest contributions to $J_J$ arise near the antinodal points of the
(bare) Fermi surface. These contributions are drastically reduced (note the logarithmic intensity scale)
with increasing stripe modulation,
as a result of the Fermi-surface distortions\cite{millis07,WV08} accompanying the stripe order.
For sizeable $\delta\Delta$, the combination of Fermi-surface and order-parameter reconstruction
even generates regions in momentum space with $J_J(\vec k)<0$,
Fig.~\ref{fig:Cstripes_J_plusminus}b.
(For $\delta\Delta=0$, this effect occurs only near the BZ diagonals due to the shift of
nodal lines from stripe order, but this has little influence on $J_J$ due to the specific momentum
dependence of the inter-layer tunneling.)
The imposed pairing modulation $\delta\Delta$ is seen to contribute to the
reduction of $J_J$, Fig.~\ref{fig:coupling_CDWdiffratios_SDW}a.

We now turn to the influence of magnetic SDW order as in Fig.~\ref{fig:cartoon}b.
As shown in Fig.~\ref{fig:coupling_CDWdiffratios_SDW}b,
SDW order alone (with CDW being parasitic only) leads to a moderate suppression of $J_J$.
Similarly, SDW order in combination with a CDW suppresses $J_J$ further
compared to the non-magnetic case,
mainly because of the additional Fermi-surface reconstruction arising from the SDW wavevector.
Note that the {\em relative} spin orientation between the two layers does not enter the
result.

Finally, we link our findings to the experimental situation.
Unfortunately, the magnitude of the charge modulation in cuprate stripes is not well known:
From resonant soft x-ray scattering\cite{abbamonte2005} on \lbcoo\ the modulation on the
oxygens was concluded to be of order factor 4, but the quantitative analysis
is model dependent. From the STM data\cite{kohsaka07} on \bscco\ and \ccoc\ one may infer a
typical modulation amplitude in the charge sector of $\pm 20 \ldots 30\%$.
In our calculation, a modulation of $\pm25\%$ in the bond kinetic energies is obtained from
$\delta t=0.023$\,eV, which gives a 20\% reduction of $J_J$ (Fig.~\ref{fig:Jcoupling_condensate}b),
while $\delta t=0.07$\,eV with a factor 90 reduction of $J_J$
corresponds to a kinetic-energy modulation of about a factor 10
(here, the neglect of Mott physics makes this number less meaningful, as the
quasiparticle theory does not account for half filling being special).

The magnitude of the magnetic order in striped 214 cuprates is known reasonably
well, at $x=1/8$ the maximum moment size is $50\ldots60\%$ of that of the undoped
compound.\cite{nachumi98,stripe_rev2}
In our mean-field calculation, $\delta\mu^\sigma\approx\pm0.07$\,eV
corresponds to a maximum moment of $0.36 \mu_B$ and gives an additional factor 2 suppression
of $J_J$.

As stripes are particularly stable near doping 1/8, it is conceivable that
\lbcoo\ has rather strong modulation in the charge sector, in which case
$J_J$ would be suppressed drastically by stripe order. This in turn would be consistent with the
fact that the 2d fluctuating pairing regime is restricted to the vicinity of $x=1/8$.
Of course, corrections beyond the mean-field picture arising from phase fluctuations,
quenched disorder, short-range magnetism, and other pseudogap physics is of importance to
fully understand the hierarchy of energy scales in \lbcoo.


\section{Conclusions}

Using a mean-field quasiparticle framework, we have calculated the inter-layer Josephson
coupling in superconducting stripe states.
We have assumed that the in-plane stripe orientations alternate from
layer to layer, as is the case in 214 cuprates with LTT lattice structure.
For realistic stripe modulation strengths, we find that the inter-layer coupling can be
easily reduced by an order of magnitude.
The primary cause of this reduction is the momentum-space mismatch between the reconstructed Fermi
surfaces of adjacent layers, while the depletion of the zero-momentum superconducting condensate
(in favor of a modulated one) is secondary.
Whether this effect is sufficient to explain the unusual properties of \lbcoo\ is not
yet clear.


\begin{acknowledgments}
Discussions with E. Fradkin, M. H\"ucker, S. Kivelson, D. Scalapino, and J. Tranquada
are gratefully acknowledged.
This research was supported by the DFG through the SFB 608, FG 538, and FG 960.
\end{acknowledgments}


\end{document}